# Marginalized Two Part Models for Generalized Gamma Family of Distributions

**Delia C. Voronca[1], Mulugeta Gebregziabher[1], Valerie L. Durkalski[1], Lei Liu[3] , Leonard E. Egede[2]**

[1]Department of Public Health Sciences, Medical University of South Carolina, 135 Cannon Street, Suite 303, Charleston, SC 29425-8350, U.S.A.

[2]Center for Health Disparities Research, Medical University of South Carolina, 135 Rutledge Avenue, Room 280H, PO Box 250593, Charleston, SC 29425, U.S.A.

[3]Department of Preventive Medicine, Northwestern University at Chicago, 680 N. Lake Shore Drive, Suite 1400, Chicago, IL 60611, U.S.A.

Correspondence to: Delia Voronca; address: 35 Cross Creek Dr., Ap. Q6, Charleston, SC; 29412; email: voronca@musc.edu; tel: 843 506 0483;

**Short title: Marginalized Two Part Models**

**Word count: abstract=186, body=6791, tables=4, figures=2, references=31**

**Summary**

Positive continuous outcomes with a point mass at zero are prevalent in biomedical research. To model the point mass at zero and to provide marginalized covariate effect estimates, marginalized two part models (MTP) have been developed for outcomes with lognormal and log skew normal distributions. In this paper, we propose MTP models for outcomes from a generalized gamma (GG) family of distributions. In the proposed MTP-GG model, the conditional mean from a two-part model with a three-parameter GG distribution is parameterized to provide regression coefficients that have marginal interpretation. MTP-gamma and MTP-Weibull are developed as special cases of MTP-GG. We derive marginal covariate effect estimators from each model and through simulations assess their finite sample operating characteristics in terms of bias, standard errors, 95% coverage, and rate of convergence. We illustrate the models using data sets from The Medical Expenditure Survey (MEPS) and from a randomized trial of addictive disorders and we provide SAS code for implementation. The simulation results show that when the response distribution is unknown or mis-specified, which is usually the case in real data sets, the MTP-GG is preferable over other models.

## 1. Introduction

Positive continuous outcomes with a point mass at zero are prevalent in biomedical research. Some common examples are antibody concentration [1], disability levels [2], medical costs [3], and individual alcohol consumption [4, 5]. Common practices for modeling these types of data are: 1) to assume a regular parametric response distribution, such as Gaussian, and fit a general linear model (GLM); 2) to exclude the zeroes and model the positive outcomes via GLM; 3) to log transform the outcome after adding a small constant to the zero values and fit a GLM; or 4) to fit a two-part (TP) model. However, a simple parametric distribution, such as Gaussian, for example, cannot accommodate the large number of zeroes nor the skewness of the data, and will therefore lead to biased inferences [6]. Similarly, a log transformation cannot normalize the data due to the spike at zero, and will often result in an asymmetric distribution [7]. Tobin, and later Heckman, acknowledged the bias from an ordinary least square regression when used for continuous data with an abundance of zeroes, and proposed solutions such as the Tobit model [8] and Heckman's selection model [9]. These models assume that the dependent variable follows a censored normal distribution and all zeroes are artificial zeroes (i.e., values below a detection limit).

Two-part models, on the other hand, assume that all zeroes are valid responses, which makes them a better alternative for skewed data with a true point mass at zero [10] resulting in a more meaningful interpretation in applications where the zeroes are legitimate responses and not simply proxies for censored data.

TP models were first suggested as a generalization of the delta distribution [11], and subsequently extended to a regression setting by including covariates in both parts of the model [7, 10, 12]. The usual terms used for the two parts are binary or occurrence and intensity or continuous [13]. In general, the binary part uses a logit or a probit link to model the probability of being a positive value, whereas the continuous part models the non-zero values. Common distributions used for the non-zero values are lognormal, gamma [10, 14], generalized gamma (GG) [15], log skew normal (LSN) [16], and normal after the Box Cox transformation [5].

One caveat with the TP model is that it leads to a conditional interpretation of the regression coefficients for the continuous part (i.e., conditional that the observed value is non-zero). In order to address this limitation, alternative methods have been proposed, such as the marginalized two part (MTP) model [13] which parameterizes the marginal mean among the zero and nonzero values directly in terms of regression coefficients. This leads to a straightforward interpretation of covariate effects on the marginal mean for the entire population comprising both the zero and nonzero responses.

The MTP was originally developed for lognormal or log skew normal distributions. In this paper, we further extend the MTP model to the more flexible generalized gamma (GG) family of distributions. The GG was first mentioned by Amoroso in 1925 [17] but was later refined by Stacy, in 1962 [15] to include 3 parameters characterizing the location, scale and shape of the distribution. Because of this flexibility, the GG has the ability to represent many different types of distributions and can model a variety of data sets with different degrees of skewness and asymmetries. The GG family includes gamma, and Weibull as subfamilies, and the lognormal as a limiting distribution [18]. All of the aforementioned distributions can be used for the continuous part of TP and MTP model and more importantly, they can lead to substantially different inferences [10] [19]. Although distributions form the GG family have been used extensively in the continuous part of standard TP models, the MTP models have not yet been extended to accommodate this comprehensive family of distributions. Our aim, is to develop

MTP models for the GG family and determine which of these is most appropriate for a real data set or whether it leads to improved performance over existing MTP models when the underlying true distribution of the data is unknown.

Our work is partly motivated by a randomized controlled trial that investigates the effect of Group Motivational Interviewing (GMI) compared to a treatment control condition (TCC) among 118 veterans. The investigator was interested to assess the effect of GMI on multiple substance use and treatment engagement outcomes [20, 21] such as peak standard ethanol content (peakSEC) or the SEC on the day that the participant drank the most and US dollars spent on alcohol (ALCmoney). The percentage of zeroes for these semi continuous outcomes was 40% and 48%, respectively. We also consider a second data example from MEPS, household component 2011 (HC-147), with an increased sample size (N=13,239) and outcomes (total health care expenditures, total emergency room expenditures) with different percentages of zeroes (16%, and 88%). The aim for this data example was to determine if there were significant differences between non-Hispanic Asians and non-Hispanic whites in terms of medical costs in 2011, without conditioning on observing a non-zero cost.

The remainder of this paper is organized as follow: In section 2 we develop MTP models for GG family of distributions (gamma, and Weibull distributions); in section 3 we compare the efficiency of different MTP models in a simulation study and present a model selection approach that uses both statistical information criteria and the estimated value of the shape and scale parameter from the MTP-GG model; in section 4 we apply our findings in two real data sets; and section 5 summarizes our findings and provides suggestions for future work. We also provide SAS code for implementation in the Appendix.

## 2. Statistical Model and Inference

In the notations used for this section, $y_i$ represents the positive continuous outcome with a point mass at zero for observation $i$, $f$ is the probability density function (pdf) corresponding to a continuous distribution defined on a positive domain, $\boldsymbol{x}_i'$ is the covariate vector corresponding to subject $i$ used for the binary part, $\boldsymbol{z}_i'$ is the covariate vector corresponding to observation $i$ used for the continuous part of a TP or MTP model, $\boldsymbol{\alpha}$ is the vector of model coefficients corresponding to the binary part of the TP or MTP model, $\boldsymbol{\delta}$ is the vector of conditional coefficients corresponding to the continuous part of a TP model and $\boldsymbol{\beta}$ is the vector of marginal coefficients corresponding to the continuous part of an MTP model. Below, we provide brief description of GG followed by descriptions of the TP and MTP models.

**2.1 The generalized gamma (GG) family of distributions**

Following the parameterization in Manning et al. [22], the pdf of a continuous random variable $Y$ following a GG distribution can be written as:

$$f(\mathrm{y};k,\mu,\sigma)=\frac{\eta^{\eta}}{\sigma y\Gamma(\eta)\sqrt{\eta}}\exp\{\mathrm{u}\sqrt{\eta}-\eta\exp(|k|\mathrm{u})\}, \quad (1)$$

where $\Gamma(.)$ is the standard gamma function, $u=sign\ (k)\ (log\ y\ -\ \mu)/\sigma$, for shape parameter $k$, location parameter $\mu>0$ and scale parameter $\sigma>0$ and $\eta=|k|^{-2}>0$.

The $s^{th}$ moment of the GG distribution with the above parameterizations is given by:

$$E\left(Y^{s}\right)=\exp\left\{\mu^{s}+\frac{s\sigma\log(\mathrm{k}^{2})}{k}+\log[\Gamma(\frac{1}{k^{2}}+\frac{s\sigma}{k})]-\log[\Gamma(\frac{1}{k^{2}})]\right\} \quad (2)$$

It follows that the mean and variance of $Y$ are given by:

$$E(Y) = \exp\{\mu + C(\sigma, k)\}, \quad (3)$$

$$\mathrm{Var}(Y) = \{\exp(\mu)\,\mathrm{k}^{2\sigma/k}\}^2 \left\{ \frac{\Gamma(\frac{1}{k^2} + \frac{2\sigma}{k})}{\Gamma(\frac{1}{k^2})} - \left[ \frac{\Gamma(\frac{1}{k^2} + \frac{2\sigma}{k})}{\Gamma(\frac{1}{k^2})} \right]^{-2} \right\} \quad (4)$$

where:

$$C(\sigma, k) = \frac{\sigma \log(\mathrm{k}^2)}{k} + \log[\Gamma(\frac{1}{k^2} + \frac{\sigma}{k})] - \log[\Gamma(\frac{1}{k^2})] \quad (5)$$

Special cases: the gamma distribution is obtained when $\sigma = k$; the Weibull distribution is obtained when $k = 1$; the limiting distribution of the GG as $k \to 0$ is the lognormal distribution.

**2.2 Two part models (TP)**

The general form of the pdf of a TP model ( $g_{TP}$ ) [7] is given by:

$$\mathrm{g}_{TP}(\mathrm{y}_i) = \begin{cases} 1 - \pi_i, \; if\; y_i = 0 \\ \pi_i f\left(y_i; \boldsymbol{x}_i{}'\boldsymbol{\delta}\right), \; if\; y_i > 0 \end{cases} \quad (6)$$

where the probability of being non-zero, $\pi_i$ , can be modeled using a logit link:

$$logit(\pi_i) = \boldsymbol{z}_i'\boldsymbol{\alpha} \quad (7)$$

The location parameter $\mu_i$ is modeled in the second part of the TP model assuming a log link:

$$\log(\mu_i) = \boldsymbol{x}_i{}'\boldsymbol{\delta} \quad (8)$$

In general, the TP model will lead to a marginal interpretation (i.e. for the entire population including zero and non-zero outcomes) of model coefficients $\boldsymbol{\alpha}$ in the binary part and a conditional interpretation of the coefficients $\boldsymbol{\delta}$ in the continuous part except for some special cases. For example, if the covariates used in the binary part are different than those used in the continuous part (no common covariate) or the zero part does not have any covariate, then the covariate regression coefficient from a TP have marginal interpretation. However, when there common covariates, the interpretation of the regression coefficient from the continuous part in a TP is conditional upon observing a non-zero outcome.

The marginal mean and variance of $Y_i$ from a TP model can be derived as:

$$\mathrm{E}\left(Y_i\right)=\pi_i \mathrm{E}\left(Y_i|Y_i>0\right),\ \mathrm{V}(Y_i)=\pi_i\left[\mathrm{E}(Y_i^2\mid Y_i>0)-\pi_i\,\mathrm{E}(Y_i\mid Y_i>0)^2\right] \tag{9}$$

For example, when GG is assumed in the continuous part, then the marginal mean is,

$$\mathrm{E}(Y_i)=\pi_i \exp\{\mu_i+C\left(\sigma,k\right)\}=\frac{1}{1+\exp\left(\boldsymbol{-z}_i'\boldsymbol{\alpha}\right)}\exp\left\{\boldsymbol{x}_i'\boldsymbol{\delta}+C\left(\sigma,k\right)\right\} \tag{10}$$

The variance of $Y_i$ corresponding to the TP model can be obtained using the variance formula in equations (9) and the $s^{\text{th}}$ moment for GG in equation (2).

### 2.3 Marginalized two part models (MTP)

The general form of the pdf for an MTP model ( $g_{MTP}$ ) [13] can be written as :

$$\mathrm{g}_{MTP}\left(\mathrm{y}_i\right)=\begin{cases} 1-\pi_i,\ if\ y_i=0 \\ \pi_i f\left(y_i;\boldsymbol{x}_i'\boldsymbol{\beta}\right),\ if\ y_i>0 \end{cases} \tag{11}$$

This gives a marginal mean of the form:

$$\mathrm{E}\left(Y_i\right)=\exp(\boldsymbol{x}_i'\boldsymbol{\beta})=\xi_i \tag{12}$$

Solving for the location parameter of the GG distribution in the expression of $E(\mathrm{Y}_i)$ equation (10), we get the following parameterization:

$$\mu_i = \boldsymbol{x}_i'\boldsymbol{\beta} - \log(\pi_i) - C(\sigma, k) \tag{13}$$

We define $C$ from equation (5) for the subfamilies of the GG distribution as follows: for the MTP-gamma distribution $C = 0$, for the MTP-Weibull distribution $C(\sigma) = \log[\Gamma(1+\sigma)]$ and for the MTP-lognormal distribution $C(\sigma) = \sigma^2 / 2$. A necessary condition for the distribution used in the continuous part of an MTP is to have a finite closed-form mean that can be parameterized as in equation (13).

### 2.4 Statistical Estimation and Inference for MTP models

The parameters of the MTP model are estimated using maximum likelihood. The general form of the likelihood function for the MTP-GG model is given by:

$$L(\pi, \mu, k, \sigma \mid y) = \prod_i (1-\pi_i)^{1(\mathrm{y}_i=0)} \{\pi_i f(\mathrm{y}_i; k, \mu_i, \sigma)\}^{1(\mathrm{y}_i>0)} \tag{14}$$

Where $f$ is the pdf of GG distribution and the conditional mean $\mu_i$ is parameterized in terms of the marginal mean $\xi_i$ , as described in section 2.3. After substituting the $\mu_i$ by equation (13) and $\pi_i$ expressed in (7), we get the following likelihood form:

$$L(\alpha,\beta,\mathrm{k},\sigma \mid \mathrm{y}) = \prod_i \left(1-\frac{1}{1+\exp\left(-\boldsymbol{z}_i'\boldsymbol{\alpha}\right)}\right)^{1(\mathrm{y}_i=0)} \left\{ \frac{|k|^{-2\,|k|^{-2}}}{\left(1+\exp\left(-\boldsymbol{z}_i'\boldsymbol{\alpha}\right)\right)\sigma y_i \Gamma(|k|^{-2})\sqrt{|k|^{-2}}} \exp\left\{Q(\alpha,\beta,\sigma,k)\right\} \right\}^{1(\mathrm{y}_i>0)}$$

where

$$Q(\alpha,\beta,k,\sigma) = sign\ (k)\ \left(log\ y_i\ -\ \left(\boldsymbol{x}_i'\boldsymbol{\beta} - \log\left(\exp it(\boldsymbol{z}_i'\boldsymbol{\alpha})\right) - C\left(\sigma,k\right)\right)\right)/\,\sigma\sqrt{|k|^{-2}}$$

$$-|k|^{-2}\ \exp(k\left(log\ y_i\ -\ \left(\boldsymbol{x}_i'\boldsymbol{\beta} - \log\left(\exp it(\boldsymbol{z}_i'\boldsymbol{\alpha})\right) - C\left(\sigma,k\right)\right)\right)/\,\sigma)$$

(15)

The likelihood function for the special cases gamma and Weibull can be derived in similar ways using the corresponding probability density functions. After taking the log of the likelihood function, the derivatives with respect to each parameter are equated to zero in order to get maximum likelihood estimates. The model based asymptotic standard errors are computed using Fisher information after substituting the maximum likelihood estimates for $\alpha,\beta,\mathrm{k},\sigma$ corresponding to the MTP-GG model:

$$\mathrm{var}(\hat{\alpha},\hat{\beta},\hat{\mathrm{k}},\hat{\sigma}) = diag\{\mathrm{I}^{-1}(\hat{\alpha},\hat{\beta},\hat{\mathrm{k}},\hat{\sigma})\} \qquad (16)$$

The beta coefficients from the MTP model represent population covariate, effects and $\exp(\beta_j)$ is interpreted as the multiplicative effect on the unconditional marginal mean of $Y$ for one unit increase in $x_j$ .

The likelihood function can be maximized using SAS PROC NLMIXED. By default, the Dual Quasi Newton Rapson algorithm is used for optimization (more details in the appendix). Confidence intervals and p-values are obtained as part of the standard output in PROC NLMIXED.

## 2.5 Model selection

Popular approaches to select the best model are the likelihood ratio test for nested models and the information criteria such as Akaike information criterion (AIC) [23] and Bayesian information criterion (BIC), for non-nested models. Both AIC and BIC are based on log likelihood and they both penalize for the number of parameters being estimated. The penalty is larger for BIC and therefore BIC usually prefers more parsimonious models. The model with the lowest AIC or BIC is selected as the best model. Another alternative for model selection is the Vuong test [24]. Due to its simplicity of implementation, the Vuong test is commonly used in the context of two part models to detect zero inflation. When the models are not strictly non-nested the use of the Vuong test is not appropriate [25]. Some models presented in this article are strictly non-nested, such as MTP-Weibull and MTP-gamma; we therefore used AIC for model selection. In addition to the AIC, we used the estimates of the shape and scale parameter of the MTP-GG as indicators for the most appropriate distribution. For example, if the shape parameter $k$ is estimated as being very low (close to zero), this would suggest that an MTP-lognormal might be appropriate. Similarly, if $k$ is close to one, then MTP-Weibull would be appropriate. If $k$ and the scale parameter $\sigma$ have similar values, then MTP-gamma would be appropriate. If the parameter estimates do not match any of these scenarios and there are no large differences between AICs, then MTP-GG is considered the best alternative.

## 3. Simulation Study

The purpose of the simulation study is to: (i) investigate the bias and consistency of the estimators in finite samples, (ii) compare asymptotic standard errors, (iii) verify if the confidence interval constructed for a parameter is able to achieve the 95% nominal level of coverage, (iv) verify if a hypothesis testing procedure will attain 0.05 size/level and if so, to determine what power is possible against different alternatives to the null hypotheses. Both percent relative mean and median are used to estimate the average bias. We consider simulation scenarios for 3 different sample sizes (N=100, N=500, N=1000), 4 distributions with different shape and scale parameters (lognormal, gamma, Weibull, GG) and 3 different covariate effects ( $\beta_2 = 0, \beta_2 = -0.5, \beta_2 = -1.5$). We compare the performance of MTP-lognormal, MTP-gamma, MTP-Weibull, MTP-GG and MTP log skew normal in terms of bias, relative efficiency, 95% coverage, type 1 error rate, attained power and convergence rates. A model was considered to have converged if the standard errors for each parameter that was estimated was non-zero and non-missing and if the convergence status outputted by SAS PROC NLMIXED indicated that the model has converged.

### 3.1. Simulation design

The data set was simulated from MTP models that assumed different distribution for the continuous part with different degrees of skewness: lognormal with $\sigma^2 = 2$ and $\sigma^2 = 4$ , gamma with shape parameter 0.2, 2, and 10, Weibull with shape parameter 0.25, 2 and 10 and GG with scale 0.3 and shape 5. The shapes of the distributions used for the continuous part of the MTP are presented in Web Figure1. The true model used for simulation is the MTP model as described in section 2.3 assuming two covariates in each part:

$$\text{logit}(\pi_i) = \alpha_0 + \alpha_1 x_{i1} + \alpha_2 x_{i2} \quad (17)$$

$$\log(E(\text{Y}_i)) = \beta_0 + \beta_1 x_{i1} + \beta_2 x_{i2} \quad (18)$$

Where $\text{x}_{i1} \sim \text{N}(10, 2)$ and $x_{\text{i}2} \sim \text{Bernoulli}(0.5)$ . The parameter true values were set to $\alpha' = (4.9, -0.4, -1)$ and $\beta' = (6.3, -0.5, -1.5)$ . In order to evaluate type 1 error rate and power, we also simulated for $\beta_2 = 0$ and $\beta_2 = -0.5$, respectively. The percentage of zeroes was about 40%, similar to one of our motivating data example and other zero-heavy data sets analyzed in the literature [5, 13, 26]. Each simulation study consists of 1000 independent samples. For comparison, we also considered a limited number of simulation scenario for 20% and 60% zeroes. These results presented in the web appendix.

### 3.2 Simulation results

In Table 1, we recorded the percent relative mean bias and the asymptotic standard errors for the marginal beta coefficients after fitting MTP-GG, MTP-gamma, MTP-Weibull, MTP-lognormal and MTP log skew normal.

For small sample size (N=100) the percent relative mean bias for the intercept was extremely high for MTP-lognormal when the data set was simulated from MTP-gamma with shape parameter 0.2 (-137.75%) or from MTP-Weibull with shape parameter 0.25 (-111.59%). Also, the asymptotic standard errors were in general larger for MTP-lognormal compared to the other models. The percent relative mean bias for the intercept was moderate for MTP-Weibull (-8.14%) when the data set was simulated from MTP-gamma with shape 0.25, and similarly for the MTP-gamma (9.27%) when the data were simulated from MTP-Weibull with shape 0.25, whereas the bias for MTP-GG was lower (2.06%, 3.99%). The MTP-gamma (6.36%, 5.33%) led to slightly biased estimates for the treatment effect when the data set was MTP-Weibull with shape 0.25 or MTP-lognormal with variance 4, respectively. In terms of percent relative mean bias for the treatment effect, MTP-Weibull (0.69%, 2.97%) performed better than MTP-gamma (3.71%, 5.33%) or MTP-GG (5.41%, 3.64%) when the data set was simulated from MTP-gamma shape 0.2 or MTP-lognormal with variance 4. MTP-GG (0.62%, 0.59%) performed better than MTP lognormal (0.73%, 1.82%), MTP-gamma (0.76%, 6.36%) or MTP-Weibull (0.99%, 1.52%) when the data set was simulated from MTP-gamma shape 10 or MTP-Weibull shape 0.25, respectively. When data were simulated from an MTP-GG model, the MTP-GG resulted in slightly lower relative percent bias comparative to MTP-Weibull and MTP-gamma for all model parameters but MTP lognormal had the lowest bias in the treatment effect (0.22%). Of note, the distribution of the parameter estimates from the 1000 simulated samples was not always symmetric or normally distributed and therefore we also recorded the median relative percent bias and median standard error which are also presented in the Web Table 1. However, results based on the median were very similar with the results based on the mean relative percent bias.

For large sample size (N=1000) the intercept was highly biased for both MTP-lognormal (-148.97, -22.17%) and MTP log skew normal (-119.93%, -44.06%) when the data set was simulated from MTP-gamma with shape parameter 0.2 or from MTP-Weibull with shape parameter 0.25. Moreover, the MTP-lognormal and MTP log skew normal had the largest asymptotic standard errors when the data set was simulated from MTP-gamma or MTP-Weibull. The percent relative mean bias in the intercept was moderately upwards for MTP-Weibull (-10.18%) when the data set was simulated from MTP-gamma with shape parameter 0.2, whereas the bias was slightly upwards for the treatment effect for MTP-gamma (2.10%) when the data set was simulated from MTP-Weibull with shape 0.25. In general, MTP-GG had lower bias for the treatment effect comparative to other models form the GG family. Moreover, MTP-GG (0.24%, 0.21%, 0.33%, -0.03%) had the lowest bias for the treatment effect among all models when the data set was simulated from MTP-Weibull with shape 0.25, MTP lognormal with variance 2 or 4 or MTP-GG. The results based on the median percent relative bias were similar with the exception the MTP log skew normal which showed less bias in the intercept (Web Table 1).

The convergence rates for all models are presented in Web Table 4. The MTP log skew normal (as low as 17%) and MTP-GG (as low as 55%) had low rates of convergence when sample size was small. For large sample size, the MTP-GG rates of convergence were low (65.3%) when the data set was simulated from MTP-lognormal variance 2 and 4, respectively. For all other scenarios, the convergence was 100%. MTP log skew normal had lower rates of convergence (between 81% and 93%) for large sample size when the data set was simulated from MTP-gamma or MTP-Weibull any shape. For small samples, it is expected that more complex models such as MT-GG and MTP log skew normal to have lower convergence rates. However, for larger samples, non-convergence may be viewed as a computational issue. For large samples, MTP-GG had lower rates of convergence only when the true model was MTP-lognormal. The lognormal distribution is a limiting distribution of GG, as the shape parameter of the GG goes to zero, which may cause the model not to converge. Based on these results, the non-convergence of GG for large sample sizes could be used to select a simpler model such as MTP-lognormal. Moreover, in a real data set different initial values can be used for the model parameters until the model reaches convergence.

In Table 2 we present the 95% coverage for $\beta_2$ corresponding to the treatment effect for different simulation scenarios. For small as well as for large sample sizes, MTP-gamma had the lowest coverage (N=100: as low as 70.8%, N=1000: as low as 58.6%) when the data set was simulated from MTP-Weibull with shape 0.25 or MTP lognormal variance 2 or 4. The departures from the nominal coverage level for MTP-gamma were more severe with increased sample size. MTP-Weibull had slightly lower than nominal coverage (~87%), for large and small sample size, when the data set was simulated from MTP-lognormal with variance 2 or 4. MTP-GG had coverage that is slightly closer to the nominal 95% for all scenarios with slightly lower coverage (N=100: 89.7%, N=1000: 91.8%) when the data set was simulated from MTP-gamma shape 0.25. When the data set was simulated from MTP-gamma shape 2 and 10, or from MTP-Weibull shape 2 or 10 the coverage of MTP-GG (93.4%, 94%, 95.9%, 94.3%) was slightly better than MTP-Weibull (92.7%, 91.3%, 94.1%, 91.4). MTP-lognormal had good coverage for all scenarios but also increased standard errors comparative to other models.

In Figure 1 we compared type 1 error rates. MTP-gamma had type 1 error rates that were higher than normal (N=100: as high as .292, N=1000: as high as .450) when the data set was simulated from MTP-lognormal with variance 2 or 4, or MTP-Weibull with shape 0.25.Type one error rates for MTP-gamma increased with large sample size. MTP-Weibull had relatively high type one error rates (N=100: as high as .122, N=1000: as high as .133) when the data set was simulated from MTP-lognormal variance 2 or 4, and the results did not improve with larger sample size. However, MTP-Weibull had nominal levels for type one error rate for all other simulation scenarios. All other MTP models had type one error rates around the 0.05 nominal level.

Power is presented in Figure 2. We compared the power levels only for scenarios where the type one error rate did not exceed nominal levels since the power may seem higher as a consequence of increased type one error rate. MTP-lognormal followed by MTP log skew normal had lowest power in detecting a treatment effect when the data set was simulated from MTP-gamma shape 0.2 and shape 2 or MTP-Weibull shape 0.25 or 2. . As expected, the power was higher for all models as the sample increased. When the data set was simulated from MTP-Weibull or from MTP-gamma shape 2 or shape 10, MTP-Weibull followed by MTP-GG had best power.

For each simulation scenario, we also recorded the average and median estimates for the shape and scale parameter after fitting MTP-GG for the 1000 simulated samples. The results are presented in Web Table 6. When the data set was simulate from MTP-gamma the shape and scale parameters of MTP-GG were very close. When the data set was simulated from MTP-Weibull, the shape parameter of MTP-GG was close to one. When the data set was simulated from MTP-lognormal the shape parameter of the MTP-GG was close to zero. All these suggest that the parameter estimates of MTP-GG could be used as an additional tool when selecting the best model fit. This can be useful especially for small sample size when the choice of the model is more important to get unbiased inference. In order to determine the best model, we also looked at the percent of times the model was selected based on the smallest AIC. For small sample sizes, AIC did not discriminate well between the models when the data set was simulated from MTP-gamma or MTP-Weibull. AIC was more efficient to identify the true model when the data set was simulated from MTP-lognormal (~78% of the time). As the sample size increased, AIC performed better and the right model was identified 78% of the times or more for all simulation scenarios. We present AIC results in Web Table 5.

Results for 20% and 60% zeroes are presented in Web Table 7, for data simulated from MTP-GG (0.3, 5) for a sample size of 1000. The results are similar to the simulation results based on 40% zeroes, suggesting that MTP-GG is a good alternative for any scenario. For 20% zeroes, the measures of performance improve for all models with the exception of type one error rates which are slightly increased comparative to 60% zeroes.

To sum up, for small sample sizes it is not straightforward to choose a model that performs well for any distribution of the data. If the sample is small and the interest is focused on estimating the treatment effect, the MTP lognormal, in general, leads to smaller bias, reduced type one error rate and good 95% coverage comparative to the other models. However, for larger sample sizes, our simulations show that the bias for the treatment effect is small and the power is optimal for the more complex models such as MTP-GG and MTP log skew normal. If the interest is prediction, the investigator should be careful when using an MTP lognormal or MTP log skew normal which lead to highly biased intercepts when the true model is MTP-gamma or MTP-Weibull with small shape parameters. In addition, the MTP-gamma and MTP-Weibull lead to increased type one error rates, regardless of sample size, when the true model is MTP lognormal. However, in general MTP Weibull performs better than MTP gamma. Therefore, we conclude that MTP-GG is a good alternative for analysis of semi-continuous data when sample size is large whereas MTP-LN is a good starting point when sample size is small. The estimated shape and scale parameters of MTP-GG can be used to identify a better model.

## 4. Data Example

### 4.1 Description of the data

The first data set is derived from a study that investigated the relative effectiveness of motivational interviewing (GMI), compared to a treatment control condition (TCC) for dually diagnosed veterans for enhancing treatment engagement and lowering substance use in an outpatient substance abuse treatment program. Participants were randomized to GMI (n=59) or TCC (n=59). Patients attended four 75- minute sessions of GMI or TCC across four consecutive days. A study coordinator blind to study conditions evaluated participants at baseline and at a 1 and 3-month follow-up. Two of the primary outcomes for the study were peak standard ethanol content (peakSEC) and money spent on alcohol in US dollars (ALCmoney), measured at 3 months follow-up. Secondary measures included Treatment Motivation Questionnaire (TMQ; [27]), the Short Inventory of Problems (SIP-R; [28]), and post-traumatic stress disorder (PTSD). The primary hypotheses were based on comparing GMI to TCC at the three month follow-up after adjusting for any baseline differences. The aim of the study was to know the impact of GMI for the overall population without restricting the interpretation only for those with positive values of the outcomes.

The second data example makes use of the publically available micro-data from the consolidated Medical Expenditure Panel Survey (MEPS), household component 2011 (HC-147). The interest is to compare the adult non-Hispanic whites with one of the fastest growing minority group in US, which is the non-Hispanic Asian population [29]. To identify the target populations, we make use of the race and ethnicity variables in MEPS-HC files that are provided to facilitated analysis for minority groups. Our aim is to determine if there are any differences in total health care expenditures (TOTEXP11), and emergency room (ER) expenditures (ERTEXP11) in 2011, between the adult (>=18 years) US noninstitutionalized non-Hispanic white and non-Hispanic Asian civilians, without conditioning on observing a non-zero expense. Of note, our method does not currently provide a way to adjust for sampling weights and therefore our analysis will produce sample specific estimates. However, our purpose is to provide a data example for the proposed methodology and therefore the adjustment for sampling weights can be disregarded in the context of this article but will be considered for future work. Results are presented on the main dependent covariate across different MTP models.

### 4.2 Results of the data analysis

#### 4.2.1 Case study 1

For the randomized clinical trial example, all secondary measures TMQ, PTSD, and SIP were included in the continuous part of the MTP model in addition to the baseline measure specific to each outcome.  Only TMQ was included in the binary part. These decisions were made based on model fit (smallest AIC). peakSEC and alcMoney have an increased percentage of zero (40% and 48%) and long far right tails with different degrees of skewness, as presented in the Web Figure 2 in the web appendix. In Web Table 4, we recorded the AIC and the log likelihood for different MTP models for each outcome (MTP-GG, MTP-gamma, MTP-Weibull, MTP lognormal, MTP log skew normal). In addition, we provided the shape and scale parameter estimated by MTP-GG model to aid in the decision of model selection, when available. For peakSEC the smallest AIC was for MTP-gamma followed by MTP log skew normal. However, the model fit were almost the same for all models with differences in AICs of one or two units. The shape and scale parameters from the MTP-GG model were close in value (0.71 and 0.54) which also suggested that true model could be MTP-gamma. Assuming MTP-gamma is the best model, we conclude no significant treatment effect for peakSEC at 3 months. MTP-GG led to similar results as MTP-gamma. For money spent on alcohol the smallest AIC was for MTP-lognormal followed by MTP-gamma. The models are not nested so we cannot use a likelihood ratio test. However, the shape parameter $k$ estimated by MTP-GG was very small (0.08) which suggested that the true model could be MTP-lognormal. Assuming MTP-lognormal is the true model, we conclude there is no treatment effect on ALCmoney at three months follow up. We also note that both MTP-gamma and MTP-Weibull resulted in significant treatment effect which could be due to a type-I error as observed in our simulation studies. MTP-GG yield similar results to best model choice, MTP lognormal.

In terms of other covariates included in the models, TMQ was negatively associated with peakSEC in the continuous part which suggests that higher levels of treatment motivation were associated with lower levels of peakSEC at 3 months follow up, PTSD was positively associated with peakSEC which suggests that having PTSD is associated with an increase in the amount of money spend on alcohol at 3 months follow-up. In addition, TMQ was significantly and negatively associated with the probability of having a non-negative outcome for both outcomes. This suggests that higher levels of motivations increase the probability of no substance use as well as the probability of spending no money on alcohol. More specific, the exponentiated alpha coefficient corresponding to TMQ, in the MTP-lognormal model for peakSEC can be interpreted as a unit increase in TMQ decreases the chances of having a non-zero outcome at 3 months by a 4% (OR=0.96; CI: 0.95 to 0.99). Similarly, the exponentiated beta coefficients can be interpreted as the percent change in the marginal mean of peakSEC for one unit increase in the covariate. For example, the subjects in the GMI treatment group had RR=0.79 (CI: 0.54 to 1.15) times lower levels of peakSEC at 3 months follow up (or a reduction of 21%) compared to subjects in the TCC group after adjusting for baseline peakSEC levels and other covariates.

In summary, the MTP models with the best model fit suggested a decrease in the substance use levels and money spent on alcohol at three months in the GMI treatment group compared to TCC group, but none reached statistical significance.

**4.2.2 Case study 2**

Summary statistics for the second data example from MEPS are presented in Appendix Table 1. The total sample size considered for analysis was 13,239, where 86.2% were whites and 13.8% were Asians. The mean age for this sample was 48.3 ($\pm$ 18.1 years) with the majority of subjects being insured (82.6%) and currently married (56%). The most prevalent comorbidities were hypertension (33.03%), joint pain (35.9%) and high cholesterol (31.75%).The outcomes considered for analysis, total health care cost and ER cost, had an increased percentage of zeroes, 16% and 88%, respectively. The mean total health care cost in US dollars was 5,124.52 ($\pm$23,472.74), whereas the mean ER cost was 178.07$\pm$ 931.44. In order to analyze these semi-continuous outcomes and obtain coefficients with marginal interpretation (i.e. unconditional on observing a positive value), we use the proposed MTP models. Based on clinician suggestion and on the literature [30], the same set of covariates is included in both parts of the model. In addition to the main dependent covariate race (Asian, white), the analysis is adjusted for important confounders, such as age, gender, marital status, education, income as percent of poverty line, insured, region, depression, cardio vascular disease, (any of myocardial infraction, angina, coronary heart disease, and other heart disease), and comorbidities such as hypertension, joint pain, stroke, emphysema, high cholesterol, arthritis, asthma, and diabetes (see Table 1). For total health care expenditures, the smallest AIC corresponds to MTP log skew normal (AIC=189,771), closely followed by MTP-GG (AIC=189,779). The parameter estimates and standard errors are shown in Table 1. Based MTP-log skew normal model, we conclude that there is a significant difference in total health care cost in 2011 between the non-Hispanic whites and non-Hispanic Asians living in US. The rate ratio of total health care cost for Asians versus total health care cost for whites was RR=exp( -0.53)=0.59 (95% CI: 0.54, 0.65) which suggests that on average, the total health care cost in 2011 for the non-Hispanic Asian population was 41% less than the total cost for the non-Hispanic white population. Of note, other models such MTP gamma (exp(-0.32)=0.73) and MTP Weibull (exp(-0.39)=.68) yield lower magnitude of associations between race and total health care expenditures. For the second outcome, total ER expenditures, the best model fit corresponds to MTP-GG (AIC=25,254), followed by MTP-log skew normal (AIC=25,276) with a difference in AICs of 22 units. The MTP-GG shape and scale parameters ( $k = 0.62,\ \sigma = 1.11$ ) do not match any subfamily of the GG distribution which also confirms that a more comprehensive model such as MTP-GG is a better model fit. Base on MTP-GG, we conclude that there is a significant difference in total ER expenditures for 2011 between whites and Asians living in US. More specific, the total ER cost for 2011 was 60% (RR=0.40, 95%CI: (0.28, 0.57)) lower for non-Hispanic Asians comparative to non-Hispanic whites living in US. For the total ER expenditures outcome, all models yield similar parameter estimation and statistical significance of the race covariate. In addition to the marginal coefficients in the continuous part, the investigator has the information on the alpha coefficients from the binary part of the MTP model which are also presented in Table1. Based on the models with the smallest AIC, the log odds of having health care expenditures (alpha=-0.73; 95% CI: 1.00, -0.57) as well as the log odds of having ER expenditures (alpha=-0.75; 95%CI: -0.99, -0.51) in 2011 were significantly lower for non-Hispanic Asians comparative to non-Hispanic whites.

## 5. Conclusions and future work

In this paper we developed marginalized two-part models for the generalized gamma family of distributions (gamma and Weibull distributions). Other distributions such as lognormal, inverse gamma, exponential, chi-square, half normal, Rayleigh, and Maxwell-Boltzmann distribution [18, 31] can also be derived as special cases of the GG.

When the sample is small, a simpler model, such as the MTP lognormal may be sufficient to obtain optimal estimation of marginal treatment effect. We also showed that for multiple simulation scenarios MTP-Weibull performed better than MTP-gamma and therefore MTP-Weibull is another good alternative for a real data application with small sample size. These findings are also consistent with other studies available in the literature [21].

Our simulation results as well as the data example, also demonstrate that when the sample is large enough, a more complex model, such as MTP-GG, will lead to more accurate and reliable results. Based on our simulations, MTP-GG is superior to all other models when data is truly MTP-GG. Moreover, MTP-GG performs better than the other models for multiple other scenarios such as when the true data is MTP-Weibull or MTP lognormal. Therefore, MTP-GG is a good alternative for marginal estimation of treatment effect and marginal predictions, regardless of the true underlying distribution of the data, and especially when the sample size is large. These results are consistent with findings from other articles [5]. One challenge with MTP-GG was convergence which is expected for data with small sample sizes since this is a more complex model. However, for larger sample sizes, non-convergence of MTP-GG, may suggest that the best fitting model is most likely MTP lognormal and therefore MTP lognormal can be used as an alternative. In a real data application, we can also improve convergence of the more complex models, by using different initial values especially for the shape and scale parameter which was an efficient way for reaching convergence in our real data example. The grid search in PROC NLMIXED could also be used to select the optimal combination of initial values for all model parameters.

Given that continuous outcomes with a point mass at zero are frequent in biomedical research, the proposed models are very useful for researchers that are interested in marginal inference for the whole population while accounting for the large proportion of zeroes. We showed that the MTP models for the GG family of distributions can accommodate a variety of data scenarios. The choice of the distribution depends on factors such as sample size, with larger sample size favoring a more complex model such as MTP GG, and a smaller sample size favoring simpler models such as MTP lognormal or MTP Weibull. Moreover, our simulation results as well as the real data applications, demonstrate that the shape/scale parameter estimates of the GG distribution can help select the best fitting model similar to what we can get from using the AIC.

For future work, it would be useful to develop a formal test based on MTP-GG parameters to decide if a simpler distribution from the GG family is more appropriate. Also, the proposed models could be further extended to longitudinal/ clustered data and their performance when missigness is present also needs to be investigated.

**Supplementary Materials**

Web appendices, tables, and figures referenced in the simulation and results sections, as well as SAS code implementing the MTP-GG, MTP-gamma, and MTP-Weibull, are available with this paper at the Statistics in Medicine website on Wiley Online Library.

**Acknowledgement**

(1) This study was partially supported by grant R01DK081121 ( PI=Egede) from the National Institute of Diabetes and Digestive and Kidney Diseases (NIDDK) and CDA-2-016-08S (PI: Santa Ana) funded by the VHA Clinical Science Research and Development (CSR&D) program. (2) The manuscript represents the views of the authors and not those of the VA, AHRQ, or CSR&D.

***Appendix***

***SAS code for MTP GG:***

```
%macro mtp_gg(data,outcome);
proc nlmixed data=&data maxiter=500;
parms  a0=0.1 a1=0.1 a2=0.1 b0=0.1 b1=0.1 b2=0.1 sigma=1  t=1;
        teta=a0+a1*x1+a2*x2;
        expteta=exp(teta);
        p=expteta/(1+expteta);
        if &outcome=0 then loglik=log(1-p);
        if &outcome>0 then do;
                mu=b0 + b1 * x1 + b2 * x2 -log(p) - log(t**(-2*sigma*t)) -
log(gamma(t*t+sigma*t)/gamma(t*t));
                eta=abs(t)**(2);
                u=sign(t)*(log(&outcome)-mu)/sigma;
                value1=(eta-.5)*log(eta) - log(sigma)-log(&outcome)-lgamma(eta);
                loglik=log(p)+value1 + u*sqrt(eta)-eta*exp(1/(abs(t))*u);
        end;
                model &outcome ~general(loglik);
                ods output "Parameter Estimates"=est_gg "Convergence Status"=conv_gg "Fit
Statistics"=fit_gg;
run;
%mend mtp_gg;
```

***SAS code for MTP Gamma (special case of GG):***

```
%macro mtp_g(data,outcome);

proc nlmixed data=&data maxiter=500;

parms  a0=0.1 a1=0.1 a2=0.1 b0=0.1 b1=0.1 b2=0.1 t=1;

        teta=a0+a1*x1+a2*x2;

        expteta=exp(teta);

        p=expteta/(1+expteta);

        if &outcome=0 then loglik=log(1-p);

        if &outcome>0 then do;

                mu=b0 + b1 * x1 + b2 * x2 -log(p) ;

                eta=abs(t)**(2);

                u=sign(t)*(log(&outcome)-mu)*t;

                value1=(eta-.5)*log(eta) + log(t)-log(&outcome)-lgamma(eta);

                loglik=log(p)+value1 + u*sqrt(eta)-eta*exp(1/(abs(t))*u);

        end;

                model &outcome ~general(loglik);

                ods output "Parameter Estimates"=est_g "Convergence Status"=conv_g "Fit
Statistics"=fit_g;

run;

%mend mtp_g;
```

***SAS code for MTP Weibull (special case of GG):***

```
%macro mtp_w(data,outcome);

proc nlmixed data=&data maxiter=500;

parms  a0=0.1 a1=0.1 a2=0.1 b0=0.1 b1=0.1 b2=0.1 sigma=1 ;

        teta=a0+a1*x1+a2*x2;

        expteta=exp(teta);

        p=expteta/(1+expteta);

        if &outcome=0 then loglik=log(1-p);

        if &outcome>0 then do;

                mu=b0 + b1 * x1 + b2 * x2 -log(p) - log(gamma(1+sigma));

                eta=abs(1)**(2);

                u=sign(1)*(log(&outcome)-mu)/sigma;

                value1=(eta-.5)*log(eta) - log(sigma)-log(&outcome)-lgamma(eta);

                loglik=log(p)+value1 + u*sqrt(eta)-eta*exp(1/(abs(1))*u);

        end;

                model &outcome ~general(loglik);

                ods output "Parameter Estimates"=est_w "Convergence Status"=conv_w "Fit
Statistics"=fit_w;

run;

%mend mtp_w;
```

Table 1. Simulation results: estimates of bias and standard errors for MTP parameters

| | | LN | | G | | W | | GG | | LSN | |
|---|---|---|---|---|---|---|---|---|---|---|---|
| **Sim. From** | Coef. | %rel. bias | ASE | %rel. bias | ASE | %rel. bias | ASE | %rel. bias | ASE | %rel. bias | ASE |
| **N=100** | | | | | | | | | | | |
| **G** | β0 | **-137.75** | 4.1909 | **2.31** | 1.7619 | **-8.14** | 2.4012 | **2.06** | 1.8797 | **-1.26** | 2.2159 |
| **s=0.2** | β1 | **-0.90** | 0.354 | **-0.78** | 0.1806 | **-0.69** | 0.2419 | **-1.53** | 0.1957 | **5.73** | 0.226 |
| | β2 | **1.05** | 1.3438 | **3.71** | 0.6402 | **2.50** | 0.9037 | **5.41** | 0.6518 | **-4.82** | 0.8009 |
| **G** | β0 | **-0.32** | 0.6584 | **0.47** | 0.6066 | **0.42** | 0.5876 | **0.57** | 0.979 | **0.76** | 0.6106 |
| **s=2** | β1 | **-0.22** | 0.0686 | **0.38** | 0.0637 | **-0.38** | 0.0619 | **-0.77** | 0.1026 | **-1.22** | 0.0644 |
| | β2 | **0.70** | 0.2659 | **0.93** | 0.247 | **1.08** | 0.2388 | **1.03** | 0.2846 | **0.78** | 0.2475 |
| **G** | β0 | **0.07** | 0.351 | **0.24** | 0.3458 | **0.53** | 0.3354 | **0.35** | 0.3487 | **0.31** | 0.3456 |
| **s=10** | β1 | **0.05** | 0.0388 | **-0.26** | 0.0383 | **-0.93** | 0.0373 | **-0.50** | 0.0386 | **-0.44** | 0.0384 |
| | β2 | **0.73** | 0.1613 | **0.76** | 0.1598 | **0.99** | 0.1561 | **0.62** | 0.1615 | **0.54** | 0.1596 |
| **W** | β0 | **-111.59** | 4.2618 | **9.27** | 2.2797 | **3.71** | 2.8852 | **3.99** | 2.9255 | **-0.51** | 3.1502 |
| **s=0.25** | β1 | **-1.90** | 0.361 | **1.53** | 0.2352 | **-0.03** | 0.2911 | **-0.15** | 0.2879 | **0.64** | 0.304 |
| | β2 | **1.82** | 1.3679 | **6.36** | 0.7794 | **1.52** | 1.0807 | **0.59** | 1.0625 | **-0.26** | 1.1165 |
| **W** | β0 | **-0.47** | 0.5622 | **0.26** | 0.5191 | **0.34** | 0.4811 | **0.37** | 0.479 | **0.39** | 0.5026 |
| **s=2** | β1 | **0.31** | 0.0594 | **-0.12** | 0.0552 | **-0.33** | 0.0515 | **-0.53** | 0.0513 | **-0.41** | 0.0544 |
| | β2 | **0.41** | 0.2317 | **0.41** | 0.2169 | **0.57** | 0.2033 | **0.60** | 0.2023 | **0.72** | 0.2062 |
| **W** | β0 | **-0.08** | 0.175 | **-0.01** | 0.1724 | **1.90** | 0.1533 | **1.89** | 0.1544 | **0.66** | 0.1664 |
| **s=10** | β1 | **0.29** | 0.0223 | **0.15** | 0.0221 | **-3.63** | 0.0199 | **-3.70** | 0.0201 | **-1.21** | 0.0214 |
| | β2 | **0.41** | 0.0933 | **0.45** | 0.0923 | **-0.55** | 0.0828 | **-0.43** | 0.0831 | **0.004** | 0.0899 |
| **LN** | β0 | **0.33** | 1.0618 | **0.78** | 0.9919 | **1.48** | 1.0454 | **1.58** | 1.0626 | **-0.47** | 1.0742 |
| **$\sigma^2$=2** | β1 | **1.51** | 0.1077 | **2.13** | 0.1027 | **1.81** | 0.1079 | **0.97** | 0.108 | **1.61** | 0.1071 |
| | β2 | **0.40** | 0.41 | **3.04** | 0.3763 | **2.34** | 0.4016 | **1.71** | 0.4114 | **0.53** | 0.4079 |
| **LN** | β0 | **1.07** | 1.4753 | **3.98** | 1.3266 | **6.16** | 1.4292 | **4.39** | 1.545 | **0.45** | 1.5089 |
| **$\sigma^2$=4** | β1 | **1.77** | 0.1465 | **2.63** | 0.137 | **2.12** | 0.1465 | **1.83** | 0.1511 | **2.24** | 0.1462 |
| | β2 | **0.33** | 0.5556 | **5.33** | 0.4848 | **2.97** | 0.5432 | **3.64** | 0.5757 | **0.39** | 0.5543 |
| **GG** | β0 | **-1.91** | 0.9507 | **-0.13** | 0.8302 | **-0.07** | 0.8481 | **0.32** | 0.8904 | **-0.07** | 0.9038 |
| **s=0.5** | β1 | **0.51** | 0.0969 | **1.51** | 0.0862 | **1.45** | 0.0879 | **0.34** | 0.0918 | **-0.06** | 0.0933 |
| **k=3** | β2 | **0.22** | 0.3708 | **1.53** | 0.3251 | **1.45** | 0.3322 | **0.91** | 0.3506 | **0.83** | 0.3502 |
| **N=1000** | | | | | | | | | | | |
| **G** | β0 | **-148.97** | 1.3527 | **0.31** | 0.5051 | **-10.18** | 0.7419 | **0.32** | 0.5004 | **-22.71** | 0.644 |
| **s=0.2** | β1 | **0.57** | 0.1131 | **-0.32** | 0.0515 | **-0.26** | 0.07448 | **-0.33** | 0.0509 | **1.25** | 0.0618 |
| | β2 | **0.86** | 0.4346 | **0.71** | 0.1967 | **0.82** | 0.286 | **0.65** | 0.1948 | **0.19** | 0.2363 |
| **G** | β0 | **-0.87** | 0.2071 | **-0.08** | 0.1878 | **-0.13** | 0.1815 | **-0.08** | 0.1877 | **-0.63** | 0.2011 |
| **s=2** | β1 | **0.17** | 0.0215 | **0.22** | 0.0197 | **0.23** | 0.0191 | **0.24** | 0.0197 | **0.21** | 0.0209 |
| | β2 | **0.19** | 0.0841 | **0.16** | 0.0775 | **0.18** | 0.0752 | **0.15** | 0.0774 | **0.09** | 0.0821 |
| **G** | β0 | **-0.14** | 0.108 | **-0.09** | 0.1061 | **-0.02** | 0.1024 | **-0.09** | 0.1061 | **-0.13** | 0.108 |
| **s=10** | β1 | **0.28** | 0.0119 | **0.25** | 0.0118 | **0.13** | 0.0114 | **0.24** | 0.0118 | **0.29** | 0.0119 |
| | β2 | **0.15** | 0.0500 | **0.17** | 0.0494 | **0.19** | 0.0481 | **0.16** | 0.0494 | **0.09** | 0.0500 |
| **W** | β0 | **-119.93** | 1.3555 | **0.74** | 0.6237 | **0.35** | 0.8892 | **0.44** | 0.9019 | **-44.06** | 1.0789 |
| **s=0.25** | β1 | **-0.97** | 0.1135 | **2.38** | 0.0637 | **0.14** | 0.0892 | **0.08** | 0.0892 | **-0.34** | 0.0985 |
| | β2 | **-0.26** | 0.4354 | **2.10** | 0.2344 | **0.36** | 0.3415 | **0.24** | 0.3412 | **0.59** | 0.377 |
| **W** | β0 | **-0.51** | 0.1752 | **0.08** | 0.1599 | **0.06** | 0.1471 | **0.07** | 0.147 | **-0.19** | 0.1636 |
| **s=2** | β1 | **-0.13** | 0.0184 | **-0.14** | 0.0169 | **-0.11** | 0.0157 | **-0.12** | 0.0157 | **-0.19** | 0.0173 |
| | β2 | **0.03** | 0.0728 | **0.07** | 0.0676 | **0.13** | 0.0632 | **0.12** | 0.0631 | **0.09** | 0.0688 |
| **W** | β0 | **0.01** | 0.0529 | **0.02** | 0.0519 | **0.05** | 0.0440 | **0.05** | 0.044 | **0.03** | 0.0519 |
| **s=10** | β1 | **-0.02** | 0.0067 | **-0.04** | 0.0066 | **-0.11** | 0.0059 | **-0.12** | 0.0059 | **-0.05** | 0.0066 |
| | β2 | **0.06** | 0.0283 | **0.07** | 0.0279 | **0.12** | 0.0240 | **0.12** | 0.0239 | **0.05** | 0.0278 |
| **LN** | β0 | **0.04** | 0.3334 | **-0.06** | 0.3022 | **1.00** | 0.328 | **0.61** | 0.3481 | **-0.25** | 0.3365 |
| **$\sigma^2$=2** | β1 | **0.21** | 0.0337 | **0.90** | 0.0311 | **0.62** | 0.0337 | **0.51** | 0.0347 | **0.25** | 0.0337 |
| | β2 | **0.35** | 0.1301 | **0.35** | 0.1191 | **0.35** | 0.1295 | **0.21** | 0.1331 | **0.39** | 0.1301 |
| **LN** | β0 | **0.14** | 0.4644 | **0.58** | 0.3993 | **4.89** | 0.4496 | **1.95** | 0.4949 | **-0.67** | 0.4798 |
| **$\sigma^2$=4** | β1 | **0.24** | 0.0458 | **1.67** | 0.0409 | **0.83** | 0.0458 | **0.43** | 0.0489 | **0.06** | 0.0459 |
| | β2 | **0.40** | 0.1766 | **0.43** | 0.154 | **0.39** | 0.1758 | **0.33** | 0.1854 | **0.34** | 0.1766 |
| **GG** | β0 | **-2.10** | 0.2999 | **0.18** | 0.2555 | **0.33** | 0.2639 | **0.28** | 0.2766 | **-1.16** | 0.2916 |
| **s=0.5** | β1 | **-0.42** | 0.0305 | **-0.43** | 0.0264 | **-0.42** | 0.0272 | **-0.45** | 0.0284 | **-0.60** | 0.0298 |
| **k=3** | β2 | **-0.10** | 0.1179 | **0.33** | 0.1021 | **0.07** | 0.1053 | **-0.03** | 0.1101 | **-0.12** | 0.1151 |

ASE=Asymptotic Standard Error; GG=generalized gamma, G=gamma, W=Weibull, LN=Log-Normal, LSN=Log-Skew-Normal; in the first column, for each MTP gamma and Weibull simulation scenario the true shape parameter is provided whereas for the lognormal distribution the true sigma value is provided; the parametrization of Weibull and gamma for the simulated data is consistent with the parametrizations described in the method section; all simulated and fitted models are MTP models: the beta coefficients represent the marginal coefficients corresponding to the continuous part of MTP model and more specific, the % relative mean bias for each beta was computed as 100*(true – mean)/abs(true). A negative value of the relative percent bias suggests the parameter is overestimated and a negative value suggests it is underestimated.

Table 2. Simulation Results: 95% CI coverage probability estimates (true $\beta_2 = -1.5$ )

| **Sim.from** | **N=100** | | | | | **N=1000** | | | | |
|---|---|---|---|---|---|---|---|---|---|---|
| **MTP** | ***LN*** | ***G*** | ***W*** | ***GG*** | ***LSN*** | ***LN*** | ***G*** | ***W*** | ***GG*** | ***LSN*** |
| **G (s=0.2)** | 93.6 | 93.0 | 97.4 | 89.7 | 97.7 | 94.7 | 92.3 | 96.9 | 91.8 | 92.5 |
| **G (s=2)** | 95.6 | 93.9 | 92.7 | 93.0 | 92.9 | 93.9 | 94.7 | 94.3 | 94.9 | 94.1 |
| **G (s=10)** | 94.0 | 94.0 | 91.3 | 93.4 | 93.4 | 95.0 | 95.6 | 93.3 | 95.4 | 95.2 |
| **W (s=0.25)** | 94.8 | 72.8 | 95.0 | 93.6 | 93.4 | 95.2 | 55.0 | 96.4 | 95.8 | 96.0 |
| **W (s=2)** | 94.3 | 95.9 | 94.1 | 94.1 | 93.5 | 94.7 | 96.0 | 95.4 | 95.6 | 94.3 |
| **W (s=10)** | 93.8 | 94.3 | 91.4 | 90.4 | 90.3 | 94.6 | 95.2 | 94.5 | 94.3 | 94.9 |
| **LN ($\sigma^2$=2)** | 95.7 | 81.5 | 88.6 | 95.2 | 94.0 | 95.4 | 76.3 | 87.2 | 94.3 | 95.4 |
| **LN ($\sigma^2$=4)** | 95.7 | 70.8 | 87.8 | 94.9 | 94.5 | 95.1 | 58.6 | 86.7 | 94.8 | 94.9 |
| **GG(s=0.5, k=3)** | 94.8 | 92.1 | 92.4 | 93.7 | 93.2 | 93.0 | 90.4 | 92.4 | 94.1 | 93.3 |

MTP=Marginalized two part model, GG=generalized gamma, G=gamma, W=Weibull, LN=Log-Normal, LSN=Log-Skew-Normal; s is the shape parameter and sigma is the true standard deviation on the log scale;

The true parameters of the simulated MTP models are $\alpha' = (4.9, -0.4, -1)$ and $\beta' = (6.3, -0.3, -1.5)$

Table 3. Model coefficient estimates, standard errors and statistical significance at alpha=0.05 level for the substance use data across different MTP models

| Outcome | Coefficients | MTP LN | MTP G | MTP W | MTP GG | MTP LSN |
|---|---|---|---|---|---|---|
| **Peak** | *Alphas* | | | | | |
| **SEC** | Int | 4.94(1.556) | 5.50(1.625) | 5.60(1.636) | 4.21(.013) | 5.51(1.663) |
| | TMQ | **-0.03(.012)*** | **-0.04(.012)*** | **-0.04(.012)*** | **-0.03(.002)*** | **-0.04(.012)*** |
| | *Betas* | | | | | |
| | Int | 3.66(.633) | 3.25(.655) | 3.15(.677) | 3.31(.674) | 3.34(.672) |
| | GMI | -0.29(.207) | -0.24(.191) | -0.18(.185) | -0.26(.201) | -0.24(.208) |
| | Bline | 0.01(.009) | 0.02(.009) | **0.02(.010)*** | 0.02(.010) | 0.02(.009) |
| | PTSD | **0.44(.218)*** | 0.36(.200) | 0.33(.195) | 0.37(.209) | 0.38(.205) |
| | SIP | 0.02(.011) | 0.01(.010) | 0.009(.010) | 0.01(.011) | 0.01(.011) |
| | TMQ | **-0.02(.006)*** | **-0.02(.006)*** | **-0.02(.006)*** | **-0.02(.006)*** | **-0.02(.006)*** |
| **Alcohol** | *Alphas* | | | | | |
| **Money** | Int | 4.58(1.489) | 4.55(1.479) | 4.56(1.490) | 4.65(.896) | NC |
| | TMQ | **-0.03(.011)*** | **-0.03(.011)*** | **-0.03(.011)*** | **-0.04(.007)*** | NC |
| | *Betas* | | | | | |
| | Int | 3.79(1.065) | 2.09(1.034) | 2.36(1.097) | 3.71(.675) | NC |
| | GMI | -0.41(.365) | **-0.85(.338)*** | **-0.80(.360)*** | -0.44(.352) | NC |
| | Bline | **0.006(.002)*** | **0.009(.002)*** | **0.009(.002)*** | **0.007(.002)*** | NC |
| | PTSD | 0.19(.367) | -0.10(.405) | -0.02(.421) | 0.19(.344) | NC |
| | SIP | 0.004(.020) | -0.01(.227) | -0.009(.023) | 0.003(.020) | NC |
| | TMQ | -0.006(.010) | 0.01(.009) | 0.008(.010) | -0.006(.006) | NC |

*significant at alpha level 0.05;

MTP=marginalized two part model; GG=generalized gamma, G=gamma, W=Weibull, LN=Log-Normal, LSN=Log-Skew-Normal;

NC=Non-Convergent; GMI= group motivational interview; SEC= standard ethanol content; PTSD=Post Traumatic Syndrome Disorder; SIP=Short Inventory of Problems; TMQ=Treatment Motivation Questionnaire;

For all outcomes, the covariates included in the continuous part of the model are the baseline measure, PTSD, SIP and TMQ; the binary part includes only the TMQ covariate;

Note that for more complex models such as MTP GG and MTP LSN it may be necessary to try different initial staring values especially for the shape/scale parameters in order to get convergence;

Best model for PEAKSEC is MTP gamma; Best model for alcohol money is MTP LN;

Table 4. Model coefficient estimates, standard errors and statistical significance at alpha=0.05 level for MEPS data across different MTP models

| Outcome | Coeff. | MTP LN | MTP G | MTP W | MTP GG | MTP LSN |
|---|---|---|---|---|---|---|
| **Total** | *Alphas* | | | | | |
| **health** | Int | 1.34(0.201) | 1.18(0.201) | 1.29(0.202) | 1.39(0.200) | 1.37(0.201) |
| **exp.** | Asian | **-0.73(0.077)*** | **-0.98(0.074)*** | **-0.80(0.078)*** | **-0.72(0.077)*** | **-0.73(0.077)*** |
| | *Betas* | | | | | |
| | Int | 6.71(0.100) | 7.20(0.088) | 6.99(0.098) | 6.70(0.099) | 6.69(0.099) |
| | Asian | **-0.54(0.050)*** | **-0.32(0.048)*** | **-0.39(0.050)*** | **-0.53(0.050)*** | **-0.53(0.050)*** |
| **Total** | *Alphas* | | | | | |
| **ER** | Int | -1.78(0.190) | -1.79(0.189) | -1.79(0.190) | -1.79(0.190) | -1.78(0.190) |
| **exp.** | Asian | **-0.75(0.122)*** | **-0.74(0.122)*** | **-0.75(0.122)*** | **-0.75(0.122)*** | **-0.75(0.122)*** |
| | *Betas* | | | | | |
| | Int | 5.44(0.264) | 5.49(0.246) | 5.48(0.256) | 5.43(0.260) | 5.44(0.264) |
| | Asian | **-0.93(0.182)*** | **-0.89(0.169)*** | **-0.90(0.175)*** | **-0.92(0.180)*** | **-0.94(0.182)*** |

*significant at alpha 0.05 level;

MTP=marginalized two part model; GG=generalized gamma, G=gamma, W=Weibull, LN=Log-Normal, LSN=Log-Skew-Normal;

For these analyses, the same set of covariates were included in both part of the MTP model based on clinicians suggestion: age (continuous), gender (male ,female), marital status (married, not married), education (less than high school, high school, college, graduate), income as percent of poverty line (below or near poverty line, low, middle and high income), insured (yes, no), region (Northeast, Midwest, South, West), depression (yes, no), cardio vascular disease (yes, no) (any of myocardial infraction, angina, coronary heart disease, and other heart disease), and comorbidities such as hypertension (yes, no), joint pain (yes, no), stroke (yes, no), emphysema (yes, no), high cholesterol (yes, no), arthritis (yes, no), asthma (yes, no) and diabetes (yes, no).

Best model for Total health care expenditures is MTP LSN; Best model for total ER expenditures is MTP GG;

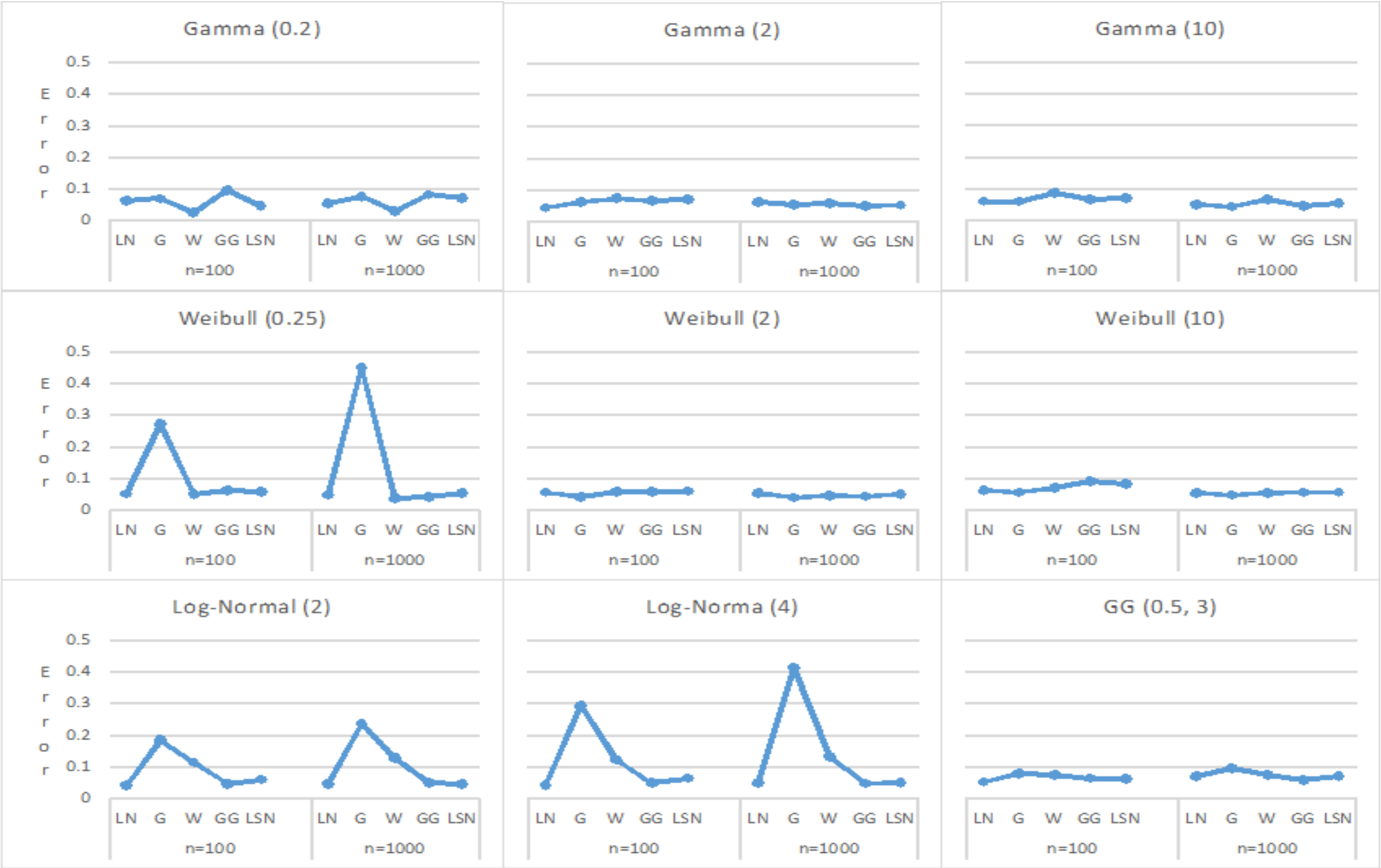

Figure 1. Type one error rate for no treatment effect: $\beta2=0$ ; GG=generalized gamma, G=gamma, W=Weibull, LN=Log-Normal, LSN=Log-Skew-Normal; n=sample size; the title of each graph depicts the distribution and the shape parameter that were used in the continuous part of the simulated MTP model; the x-axis represents the MTP model that was fitted to the simulated data;

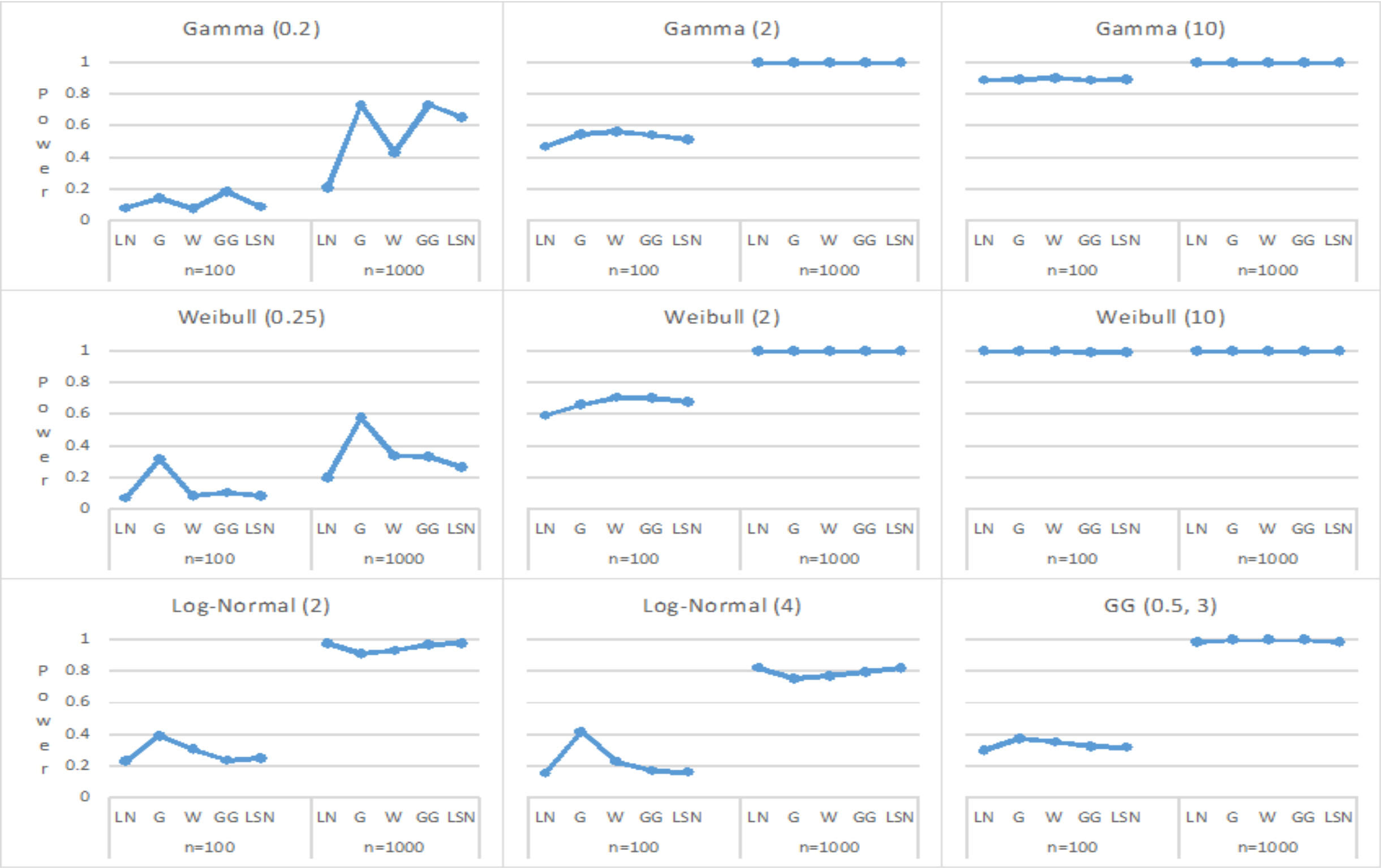

Figure 2. Power to detect a treatment effect of $\beta2=-0.5$ ; GG=generalized gamma, G=gamma, W=Weibull, LN=Log-Normal, LSN=Log-Skew-Normal; n=sample size; the title of each graph depicts the distribution and the shape parameter that were used in the continuous part of the simulated MTP model; the x-axis represents the MTP model that was fitted to the simulated data;